\documentstyle[12pt]{article}

\begin{document}
\title{The search for nonstandard Higgs bosons at FNAL and LHC}
\author{N.V.Krasnikov \thanks{E-mail address: KRASNIKO@MS2.INR.AC.RU}
\\Institute for Nuclear Research\\
60-th October Anniversary Prospect 7a,\\ Moscow 117312, Russia}
\date{September 1995}
\maketitle
\begin{abstract}
We discuss the phenomenology of many Higgs doublet model - the model where
each Higgs doublet couples with each own quark and lepton with relatively
big Yukawa coupling constants. Namely, we discuss the discovery potential
of the Higgs bosons at LHC and find lower mass limits on nonstandard
Higgs bosons from FNAl data.

\end{abstract}
\newpage

The aim of this paper is the discussion of the phenomenology of
many Higgs doublet models \cite{1}-\cite{14}. We shall restrict ourselves
to many Higgs doublet models where each Higgs doublet couples with its own
quark with relatively big Yukawa coupling constant. Many Higgs doublet models
naturally arise in some nonsupersymmetric and supersymmetic GUT models
\cite{12}-\cite{14}. Here we consider the phenomenology of many Higgs
doublet models. Namely, we discuss the discovery potential of the Higgs bosons
with relatively big Yukawa coupling constants with quarks at LHC and find
lower mass bound for nonstandard Higgs bosons from FNAL data.

To be precise in this paper we consider the model with 6 Higgs
doublets \cite{12}-\cite{14} where each Higgs doublet couples with
its own quark and lepton. The Yukawa Lagrangian of the model has the form
\begin{equation}
L_{Y} = \sum_{i,j=1}^{3} [h^{u}_{ij} \overline{A}_{Li} u_{Rj}H_{ju} +
h^{d}_{ij}\overline{A}_{Li} d_{Rj} H_{jd} + h^{l}_{ij}L_{Li}e_{Rj}H_{jd}]
+ h.c.\:,
\end{equation}

where $ A_{L1} = (u,d)_{L},\: A_{L2} = (c,s)_{L},\: A_{L3} = (t,b)_{L},\:
u_{R1} = u_{R},\: u_{R2} = c_{R},\: u_{R3} = t_{R},\: d_{R1} = d_{R},\:
d_{R2} = s_{R},\: d_{R3} = b_{R},\: H_{1u} = (H_{uu}, H_{ud}),\:
H_{2u} = (H_{cc},\: H_{cs}),\: H_{3u} = (H_{tt}, H_{tb}),\:
H_{1d} = (H_{du}, H_{dd}),\: H_{2d} = (H_{sc}, H_{ss}),\:
H_{3d} = (H_{bt},H_{bb})$

The Lagrangian (1) is invariant under the transformations
\begin{equation}
H_{ju} \rightarrow  exp(i\alpha_{j}) H_{ju}\:,
\end{equation}
\begin{equation}
H_{jd} \rightarrow exp(i\beta_{j})H_{jd}\:,
\end{equation}
\begin{equation}
u_{Rj} \rightarrow  exp(-i\alpha_{J}) u_{Rj}\:,
\end{equation}
\begin{equation}
d_{Rj} \rightarrow exp(-i\beta_{j}) d_{RJ}\:,
\end{equation}
\begin{equation}
e_{Rj} \rightarrow exp(-i\beta_{j})e_{Rj}
\end{equation}
To get rid of the problems with neutral flavor changing currents
\cite{1,2} we postulate that the matrices $h^{u}_{ij}, h^{d}_{ij}$ and
$h^{e}_{ij}$ are unitare. After the diagonalization of the mass matrices we
find that in considered model each quark interacts with its own higgs doublet .
We assume that all Yukawa coupling constants in the diagonal basis are
not small and the difference in masses between different quarks and
leptons is due to the difference in vacuum expectation values for different
Higgs doublets. The Higgs potential has the form
\begin{equation}
V = \lambda(H^{+}_{3u}H_{3u} - c^2)^2 + L_{M} + \delta L ,
\end{equation}
\begin{equation}
L_{M} = M^{2}_{2u}H^{+}_{2u}H_{2u} + M^{2}_{1u}H^{+}_{1u}H_{1u} +
M^{2}_{3d}H^{+}_{3d}H_{3d} + M^{2}_{2d}H^{+}_{2d}H_{2d} +
M^{2}_{1d}H^{+}_{1d}H_{1d}\: ,
\end{equation}
\begin{equation}
\delta L  = - (\overline{H}^{+}_{3u}[m^2_1H_{1d} + m^2_2H_{2d} +
m^2_3H_{3d} + m^2_4\overline{H}_{1u} + m^2_5\overline{H}_{2u}] + h.c.)\:,
\end{equation}
\begin{equation}
\overline{H}_{i} = \epsilon^{ij}H^{+}_{j}
\end{equation}
The term $\delta L$ violates the discrete symmetry (3) in a soft way. At
the first stage of the symmetry breaking only $H_{3u}$ Higgs doublet
acquires a nonzero vacuum expectation value that leads to the electroweak
symmetry breaking and to the generation of the t-quark mass. Due to the
soft mass term $\delta L$ after the first stage of breaking other Higgs
doublets acquire nonzero vacuum expectation values
\begin{equation}
<H_{1d}> = \frac{m^2_1}{M^2_1}<\overline{H}_{3u}>,...,
\end{equation}
For nonsmall Yukawa coupling constants the main reaction for the production
of the Higgs doublets corresponding to the first and second generations is
quark-antiquark fusion. The phenomenology of the Higgs doublet corresponding
to the third generation is very similar to the phenomenolgy of the
standard 2 Higgs doublet model and we shall not discuss it.
The cross section for the quark-antiquark fusion in quark-parton model is
given by the standard formula \cite {15}
$$
\sigma (AB \rightarrow H_{q_{i}q_{j}} +X) =
\frac{4\pi^2\Gamma(H_{q_{i}q_{j}} \rightarrow
\overline{q}_{i}q_{j})}{9 M^{3}_{H}}\int^{1}_{\frac{M^{2}_{H}}{s}}
[\overline{q}_{Ai}(x,\mu)q_{Bj}(x^{-1}M^2_Hs^{-1}, \mu)
$$
$$
+ q_{Aj}(x,\mu)\overline{q}_{Bj}(x^{-1}M^{2}_{H}s^{-1}, \mu)]
\eqno{(12)}
$$
Here $\overline{q}_{Ai}(x,\mu)$ and $q_{Aj}(x ,\mu)$ are parton distributions
of the antiquark $\overline{q}_{i}$ and quark $q_{j}$ in hadron A at the
normalization point $\mu \sim M_{H}$ and the
$\Gamma(H_{q_{i}q_{j}} \rightarrow \overline{q}_{i}q_{j})$  is the
hadronic decay width of the Higgs boson into quark-antiquark pair.
For the Yukawa Lagrangian
\begin{equation}
L_{Y} = h_{q_{i} q_{j}} \overline{q}_{Li}q_{RJ} H_{q_{i}
q_{j}} + h.c.
\end{equation}
the hadronic decay width for massless quarks is
\begin{equation}
\Gamma(H_{q_{i}q_{j}} \rightarrow \overline{q}_{i}q_{j}) =
\frac{3M_{H}h_{q_{i}q_{j}}^2}{16\pi}
\end{equation}
We have calculated the cross sections for the Higgs production using
the parton distributions of ref.\cite{16}. We have used both set1 and set2
of ref.\cite {16}. The results of our calculations  are presented in
tables 1 - 6. It should be noted that the results presened in tables 1 - 6
correspond to the Yukawa coupling constants $h^{2}_{ij} = 1$. For the case
of arbitrary Yukawa coupling $h_{ij}$ the corresponding cross section
$\sigma_{ij}$ is proportional to $h^{2}_{ij}$.
In our calculations we took the value of the renormalization
point $\mu$ equal to the mass M of the corresponding Higgs boson. We have
checked also that the variation  of the renormalization point $\mu$ in
the interval $0.5M - 2M$ leads to the variation of the cross sections less
than 50 percent. In considered model Higgs bosons couple both with quarks
and leptons so the best signature is the search for the Higgs boson decays
into lepton pairs for electrically neutral Higgses. For charged Higgses
the best way to detect them is to look for their decays into charged
leptons and neutrino. The Higgs doublets which couple with up quarks in model
with massless neutrino don't couple with leptons so the single way to
detect them is to look for the resonance type structure in the differential
dijet cross section on the dijet mass as in the case of all Higgs bosons ,
since in considered model all Higgs bosons decay mainly into
quark-antiquark pair that leads at the hadron level to the dijet events.
However the accuracy of the determination of the dijet cross section is
not very high and besides the typical accuracy in the determination of the
dijet invariant mass is $O(10)$ percent so it is difficult to find stringent
bound on the Higgs mass by the measurement of dijet differential cross section.
It should be noted that in considered many Higgs doublet model due to the
smallness of the vacuum expectation values of the Higgs doublets
corresponding to the u,d,s and c quarks after the electroweak symmetry
breaking the mass splitting inside the Higgs doublets is small, so
in such models the search for neutral Higgs boson decaying into lepton
pair is in fact the search for the Higgs isodoublet as a whole.
Consider the electrically neutral Higgs boson that couples with strange quark
and $\mu$-meson. For the s-quark mass \cite{17}
 $m_{s}(1 Gev) = (150 - 200)\:Mev$ after taking into account QCD evolution
of the running quark mass we find that
$Br(H_{ss} \rightarrow \mu^{+}\mu^{-}) =(0.3 -0.6)$. So by the measurement
of muon pairs we can look for the corresponding Higgs boson. The main
background comes from the Drell-Yan process but for relatively big value
of the  Yukawa coupling constant $h^{2} \geq O(10^{-3})$ the Drell-Yan
background is small. For the Higgs field
which couples with d-quark and electron for the d-quark mass \cite{18}
$m_{d}(1Gev) = (5-10)\:Mev$
we find that  $Br(H_{dd} \rightarrow e^{+}e^{-}) =
(0.5 - 1)\cdot 10^{-2}$, so again by the measurement of the
electron-positron invariant mass pair we can detect the Higgs bosons or to
derive the lower bound on their masses.

FNAL data \cite{18} on dilepton production agree with the predictions
of the standard Weinberg-Salam model. From our calculations
of the cross sections of the production of the Higgs bosons
(see tables 1 - 6) requiring the existence of less than 10 events with
muon antimuon or electron positron pairs for the integrated luminosity
of 70 inverse pikoborn at FNAL $p\overline{p}$ collider we find that
\begin{equation}
M_{H_{ss}} \geq 400\: Gev\: ,
\end{equation}
\begin{equation}
M_{H_{dd}} \geq 400\: Gev
\end{equation}
for $h_{ss} = h_{dd} =1$ and
\begin{equation}
M_{H_{ss}} \geq 300\: Gev\:,
\end{equation}
\begin{equation}
M_{H_{dd}} \geq 280 Gev
\end{equation}
for $h^{2}_{ss} = h^{2}_{d{d}} = 0.1$.
In fact, the strategy of the search for additional Higgs doublets with
big Yukawa couplings by the measurement of dilepton mode at supercolliders
is very similar to the strategy of the search for the additional neutral
vector bosons at supercolliders by the measurement of dilepton
mode \cite{17,18}.
Recently obtained FNAL bound $ M_{Z^{'}} \geq 650\: Gev $ \cite{18}
for new neutral $Z^{'}$-boson (in the assumption that quark and lepton coupling
constant of $Z{'}$-boson coincide with the corresponding coupling constants
of Z-boson) are slightly more stringent that our bounds (15-19).

For LHC for full integrated luminosity $\int L = 10^{5}(pb)^{-1}$ and
for $\sqrt{s} = 15\:Tev$   requiring the existence of 100 events with
lepton pairs we find that for the integrated luminosity
$L = 10^{5}(pb)^{-1}$ it would be possible to discover the Higgs
bosons $H_{ss}$ and $H_{dd}$ with
the masses up to 3000 Gev  for $h^2=1$  and with the masses
up to 2000 Gev  for $h^2=0.1$. It should be noted that the main background
comes from the Drell-Yan process however for not very small
$h^{2} \geq O(10^{-3})$ Yukawa coupling constants
and for the measurement of dilepton mass with the accuracy better than
2 percent the background is small.

To conclude, in this note we have studied the perspectives of the discovery
of the nonstandard Higgs bosons, namely Higgs bosons whith relatively
big Yukawa coupling constant with quarks at LHC. The best way to
detect such Higgs bosons is the measurement of the dilepton invariant
masses. We have found also that FNAL data lead to the bounds (15-19) on the
nonstandard Higgs bosons. LHC will be able to discover the nonstandard
Higgs bosons with masses up to 2-3 Tev.

I am indebted to the collaborators of the INR theoretical department for
discussions and critical comments. The research described in this publication
was made possible in part by Grants N6G000, N6G300 from the International
Science Foundation and by Grant 94-02-04474-a of the Russian Scientific
Foundation.

\newpage

Table 1. The cross sections $\sigma(p \overline{p} \rightarrow H_{ij}
\, + \, ...)$ in pb for different values of the Higgs boson masses
for Yukawa coupling constants $h^{2}_{ij} = 1$ and for
normalization point $\mu = M_{H}$ at FNAL (set 1).

\begin{center}
\begin{tabular}{|l|l|l|l|l|l|l|}
\hline
M(Gev) & $\sigma_{cc}$ & $\sigma_{ss}$ & $\sigma_{uu}$ & $\sigma_{dd}$
&$\sigma_{ud}$ &$\sigma_{sc}$  \\
\hline
1000 & $9.5\cdot10^{-8}$  & $5.1\cdot10^{-7}$ & 0.11 & 0.0053 & 0.025 &
$2.2\cdot10^{-7}$   \\
\hline
900 & $1.0\cdot10^{-6}$ & $6.5\cdot10^{-6}$ & 0.43 & 0.025 & 0.11 &
$2.6\cdot10^{-6}$ \\
\hline
800 & $9.5\cdot10^{-6}$ & $6.8\cdot10^{-5}$ & 1.4 & 0.11 & 0.41  &
$2.6\cdot10^{-5}$  \\
\hline
700 & $ 8.2\cdot10^{-5}$ & $6.2\cdot10^{-4}$ & 4.5 & 0.44 & 1.5 &
$2.3\cdot10^{-4}$ \\
\hline
600 & $6.9\cdot10^{-4}$ & 0.0053 & 13 & 1.7 & 4.9 & 0.0013 \\
\hline
500 & 0.0067 & 0.044 & 39 & 6.3 & 16 & 0.016 \\
\hline
400 & 0.057 & 0.39 & 120 & 24 & 55 & 0.15 \\
\hline
300 & 0.65 & 3.7 & 390 & 100 & 200 & 1.6 \\
\hline
200 & 10 & 48 & $1.6\cdot10^{3}$ & 560 & $ 0.97\cdot10^{3}$ & 22 \\
\hline
150 & 53 & 220 & $4.0\cdot10^{3}$ & $1.6\cdot10^{3}$ &  $2.6\cdot10^{3}$ & 110
\\
\hline
\end{tabular}
\end{center}

Table 2. The same for FNAL (set 2)

\begin{center}
\begin{tabular}{|l|l|l|l|l|l|l|}
\hline
M(Gev) & $\sigma_{cc}$ & $\sigma_{ss}$ & $\sigma_{uu}$ & $\sigma_{dd}$ &
$\sigma_{ud}$ & $\sigma_{sc}$ \\
\hline
1000 & $2.5\cdot10^{-6}$ & $7.7\cdot10^{-6}$ & 0.12 & 0.067 & 0.029 &
$4.4\cdot10^{-6}$ \\
\hline
900 & $2.0\cdot10^{-5}$ & $6.0\cdot10^{-5}$ & 0.43 & 0.033 & 0.12 &
$3.5\cdot10^{-5}$ \\
\hline
800 & $1.4\cdot10^{-4}$ & $4.1\cdot10^{-4}$ & 1.4 & 0.14 & 0.46 &
$2.4\cdot10^{-4}$ \\
\hline
700 & $8.4\cdot10^{-4}$ & 0.0027 & 4.4 & 0.54 & 1.6 & 0.0013 \\
\hline
600 & 0.0053 & 0.016 & 13 & 2.0 & 5.2 & 0.0093 \\
\hline
500 & 0.029 & 0.095 & 37 & 7.3 & 17 & 0.053 \\
\hline
400 & 0.19 & 0.63 & 110 & 27 & 56 & 0.35 \\
\hline
300 & 1.4 & 4.9 & 360 & 110 & 200 & 2.6 \\
\hline
200 & 16 & 54 & $1.5\cdot10^{3}$ & 580 & $0.93\cdot10^{3}$ & 28 \\
\hline
150 & 61 & 230 & $3.6\cdot10^{3}$ & $1.6\cdot10^{3}$ & $2.4\cdot10^{3}$ & 120
\\
\hline
\end{tabular}
\end{center}

\newpage

Table 3. The cross sections $\sigma(pp \rightarrow H_{ij} \,+\,...)$
in pb for different values of the Higgs boson masses
for Yukawa coupling constants $h^{2}_{ij} =1$ and for the
renormalization point $\mu = M_{H}$ at LHC (set 1, $\sqrt{s} = 10\, Tev$).

\begin{center}
\begin{tabular}{|l|l|l|l|l|l|}
\hline
M(Gev) & $\sigma_{cc}$ & $\sigma_{ss}$ & $\sigma_{uu}$ & $\sigma_{dd}$ &
$\sigma_{sc}$ \\
\hline
3000 & $3.1\cdot10^{-6}$ & $5.2\cdot10^{-4}$ & 0.040 & 0.12 & $4.0\cdot10^{-5}$
\\
\hline
2000 & $6.3\cdot10^{-4}$ & 0.027 & 0.83 & 0.32 & 0.0041 \\
\hline
1500 & 0.011 & 0.24 & 4.4 & 1.9 & 0.049 \\
\hline
1200 & 0.064 & 0.97 & 13 & 5.9 & 0.25\\
\hline
1000 & 0.24 & 2.8 & 29 & 14 & 0.81 \\
\hline
800 & 1.0 & 8.8 & 75 & 37 & 3.1 \\
\hline
600 & 5.4 & 35 & 220 & 120 & 13 \\
\hline
500 & 14 & 76 & 430 & 240 & 32 \\
\hline
400 & 41 & 190 & 940 & 540 & 88 \\
\hline
300 & 140 & 580 & $2.5\cdot10^{3}$ & $1.5\cdot10^{3}$ & 290 \\
\hline
\end{tabular}
\end{center}

Table 4. The same for LHC (set 2,  $ \sqrt{s} = 10 Tev$)

\begin{center}
\begin{tabular}{|l|l|l|l|l|l|}
\hline
M(Gev) & $\sigma_{cc}$ & $\sigma_{ss}$ & $\sigma_{uu}$ & $\sigma_{dd}$ &
$\sigma_{sc}$ \\
\hline
3000 & $2.3\cdot10{-4}$ & 0.0040 & 0.096 & 0.033 & $9.0\cdot10^{-4}$ \\
\hline
2000 & 0.080 & 0.088 & 1.3 & 0.57 & 0.027 \\
\hline
1500 & 0.060 & 0.51 & 5.7 & 2.7 & 0.18 \\
\hline
1200 & 0.24 & 1.7 & 15.3 & 7.6 & 0.63 \\
\hline
1000 & 0.68 & 4.0 & 32 & 17 & 1.7 \\
\hline
800 & 2.2 & 11 & 76 & 42 & 4.9 \\
\hline
600 & 8.7 & 37 & 220 & 120 & 18 \\
\hline
500 & 19 & 77 & 410 & 240 & 39 \\
\hline
400 & 49 & 180 & 880 & 530 & 94 \\
\hline
300 & 150 & 550 & $2.3\cdot10^{3}$ & $1.4\cdot10^{3}$ & 290 \\
\hline
\end{tabular}
\end{center}
\newpage

Table 5. The same for LHC (set 1, $\sqrt{s} = 15 Tev$)

\begin{center}
\begin{tabular}{|l|l|l|l|l|l|}
\hline
M(Gev) & $\sigma_{cc}$ & $\sigma_{ss}$ & $\sigma_{uu}$ & $\sigma_{dd}$ &
$\sigma_{sc}$ \\
\hline
4000 & $2.9\cdot10^{-7}$ & 0.00092 & 0.062 & 0.016 & 0.0054 \\
\hline
3000 & $0.70\cdot10^{-4}$ & 0.012 & 0.37 & 0.14 & 0.0036 \\
\hline
2000 & 0.0059 & 0.23 & 3.5 & 1.5 & 0.044 \\
\hline
1500 & 0.063 & 1.3 & 14 & 6.3 & 0.42 \\
\hline
1200 & 0.30 & 4.1 & 34 & 17 & 1.2 \\
\hline
1000 & 0.94 & 10 & 70 & 36 & 3.6 \\
\hline
800 & 3.4 & 27 & 160 & 86 & 8.1 \\
\hline
600 & 15 & 90 & 440 & 250 & 32 \\
\hline
500 & 35 & 180 & 830 & 490 & 71 \\
\hline
400 & 95 & 430 & $1.7\cdot106{3}$ & $1.1\cdot10^{3}$ & 180 \\
\hline
\end{tabular}
\end{center}

Table 6. The same for LHC (set 2, $\sqrt{s} = 15 Tev$)

\begin{center}
\begin{tabular}{|l|l|l|l|l|l|}
\hline
M(Gev) & $\sigma_{cc}$ & $\sigma_{ss}$ & $\sigma_{uu}$ & $\sigma_{dd}$ &
$\sigma_{sc}$ \\
\hline
4000 & $2.5\cdot10^{-4}$ & 0.0067 & 0.12 & 0.045 & 0.0013 \\
\hline
3000 & 0.0027 & 0.056 & 0.71 & 0.31 & 0.013 \\
\hline
2000 & 0.051 & 0.59 & 5.1 & 2.5 & 0.17 \\
\hline
1500 & 0.28 & 2.3 & 17 & 8.7 & 0.80 \\
\hline
1200 & 0.91 & 6.1 & 39 & 21 & 2.4 \\
\hline
1000 & 2.2 & 13 & 75 & 42 & 5.3 \\
\hline
800 & 6.3 & 31 & 160 & 95 & 14 \\
\hline
600 & 22 & 94 & 440 & 260 & 45 \\
\hline
500 & 45 & 190 & 810 & 500 & 93 \\
\hline
400 & 110 & 420 & $1.7\cdot10^{3}$ & $1.1\cdot10^{3}$ & 210 \\
\hline
\end{tabular}
\end{center}

\end{document}